\documentclass[a4paper,twocolumn,11pt,unpublished]{quantumarticle}
\pdfoutput=1
\usepackage[utf8]{inputenc}
\usepackage[english]{babel}
\usepackage[T1]{fontenc}
\usepackage{amsmath}
\usepackage{hyperref}

\usepackage{graphicx}
\usepackage{subfigure}
\usepackage{epstopdf}
\usepackage{graphicx}

\usepackage{amsfonts}
\usepackage{amssymb, color, bm}

\usepackage{currvita}

\usepackage{algorithmicx}
\usepackage[noend]{algpseudocode}
\usepackage{algorithm,setspace}

\usepackage{physics}

\usepackage{multicol}
\usepackage{lipsum}
\usepackage{mwe}

\usepackage{tikz}
\usepackage{lipsum}

\begin{document}

\title{Adaptive Bayesian Single-Shot Quantum Sensing}

\author{Ivana Nikoloska}
\affiliation{Signal Processing Systems Group,  Department of Electrical Engineering, Eindhoven University of Technology, Eindhoven, 5612 AP, The Netherlands}
\email{i.nikoloska@tue.nl}
\author{Ruud van Sloun}
\affiliation{Signal Processing Systems Group,  Department of Electrical Engineering, Eindhoven University of Technology, Eindhoven, 5612 AP, The Netherlands}
\author{Osvaldo Simeone}
\affiliation{KCLIP lab, Center for Intelligent Information Processing Systems, Department of Engineering, King's College London, Strand, London, WC2R 2LS, United Kingdom}

\maketitle

\begin{abstract}
Quantum sensing harnesses the unique properties of quantum systems to enable precision measurements of physical quantities such as time, magnetic and electric fields, acceleration, and gravitational gradients well beyond the limits of classical sensors. However, identifying suitable sensing probes and measurement schemes can be a classically intractable task, as it requires optimizing over Hilbert spaces of high dimension. In  variational quantum sensing,  a probe quantum system is generated via a parameterized quantum circuit (PQC), exposed to an unknown physical parameter through a quantum channel, and measured  to collect classical data. PQCs and measurements are typically optimized using offline strategies based on frequentist learning criteria. This paper introduces an adaptive  protocol that uses Bayesian inference to optimize the sensing policy via the maximization of the active information gain. The proposed variational methodology is tailored for non-asymptotic regimes where a single probe can be deployed in each time step, and is extended to support the fusion of estimates from multiple quantum sensing agents. 


\end{abstract}

\section{Introduction}

\subsection{Context and Motivation}

Quantum sensing is fundamentally enabled — and constrained — by the principles of quantum mechanics. Whilst these principles set intrinsic limits on measurement precision, they also introduce opportunities for enhanced sensitivity by leveraging non-classical phenomena such as quantum coherence and entanglement. Quantum sensing capitalizes on these effects to exceed the performance bounds of classical sensors, with potential applications across domains including gravitational-wave detection and biomedical imaging \cite{degen2017quantum,yu2022exposing,danilishin2012quantum,schwartz2019blueprint,budakian2024roadmap}.


Translating these quantum advantages into practical devices, however, remains a significant challenge. Real-world quantum sensors must operate on current-generation hardware, which is inherently limited by noise, decoherence, and physical constraints. In particular, devices in the noisy intermediate-scale quantum (NISQ) regime are prone to coherence loss and sampling inaccuracies, which can adversely affect measurement fidelity. 
Furthermore, crafting near-optimal quantum sensing protocols for a given metrological objective involves navigating high-dimensional Hilbert spaces, a problem that is generally computationally prohibitive using exhaustive search methods.

\begin{figure*}
    \centering
    \includegraphics[width = 0.99\textwidth]{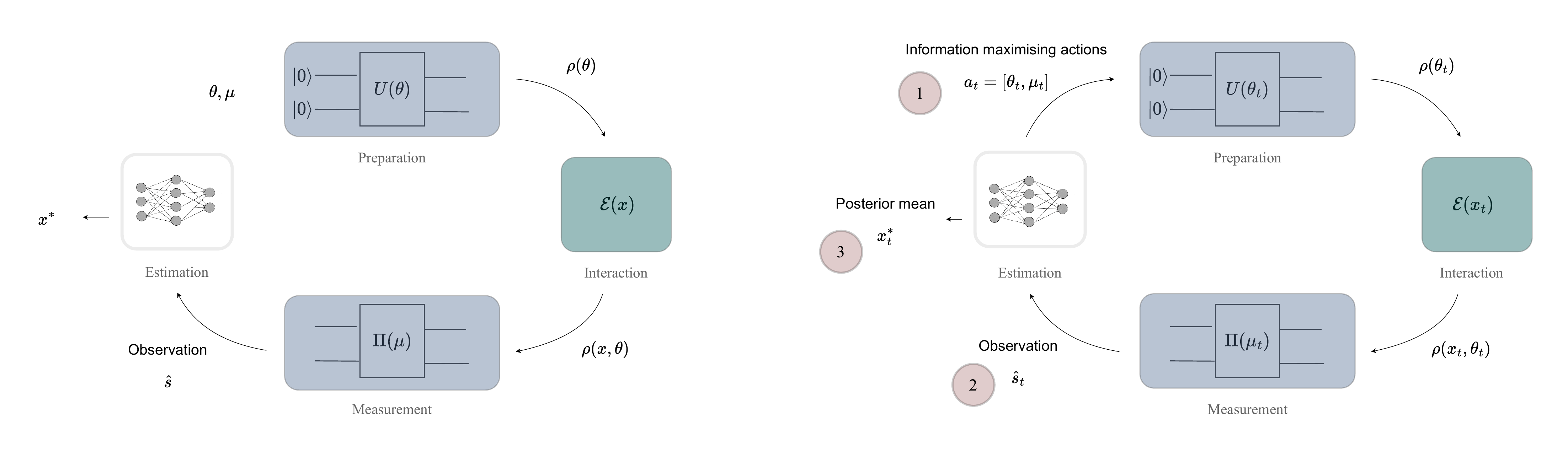}
    \caption{(left) An illustration of variational quantum sensing. A probe quantum state is generated by a $n$-qubit quantum circuit, parametrized by the fixed vector $\theta$. The probe state interacts with the parameter of interest $x$, and the perturbed state $\rho(\theta,x)$ is then measured using $\mathcal{M} = \{\Pi_m(\mu)\}$, parameterized by vector $\mu$. The parameters are fixed and multiple probes are used to obtain an estimate $x^*$.  (right) Adaptive Bayesian single-shot quantum sensing scheme under study. The probe is represented as an experimenter, using actions $a_t=[\theta_t, \mu_t]$ and pursuing an information seeing policy about the time-varying parameter of interest $x_t$. 
    }
    \label{fig:sensing_scheme_indirect}
\end{figure*}

To address this, variational quantum algorithms (VQA) \cite{schuld2021machine,simeone2022introduction} have emerged as a promising class of techniques for optimizing quantum systems within the limitations of NISQ hardware. In the realm of sensing, variational quantum sensing (VQS) frameworks \cite{meyer2021variational,maclellan2024end,nolan2021machine,xiao2022parameter,nikoloska2025dynamic} employ parameterized quantum circuits (PQCs) to design probe states and measurement observables that are  optimized for the estimation task at hand.

As shown in Fig. \ref{fig:sensing_scheme_indirect} (left) typical VQS pipeline involves generating a quantum probe state via a PQC, allowing it to interact with an unknown parameter through a quantum channel, and measuring the resulting state to obtain classical data. This data is then processed by a classical estimation routine to produce an estimate of the parameter. 

As summarized in the next subsection, most works on VQS rely on batch learning strategies that build on frequentist principles. These methods do not allow for the active adaptation of the sensing policy on the basis of prior information and measurement outcomes. This is the knowledge gap that this work is intended to address, whilst also allowing for information fusion across multiple sensors.

\subsection{Background}

Recent work has provided initial evidence for the capabilities and limitations of VQS. { As anticipated, most of this prior work relies on offline, batch optimization strategies leveraging real or simulated data \cite{nolan2021machine, xiao2022parameter}.} For instance, metrologically useful GHZ states \cite{nolan2017optimal} have been shown to be learnable by quantum graph neural networks \cite{verdon2019quantum}. A flexible variational framework for multi-parameter estimation has been developed in \cite{meyer2021variational}, whilst reference \cite{maclellan2024end} implements a complete end-to-end VQS protocol on NISQ hardware. 

The papers summarized above have adopted a frequentist formulation to the learning problem, aiming to minimize the training loss. Unlike frequentist learning, Bayesian methods have the key advantage that they inherently capture the uncertainty about unknown parameters of interest via the update of posterior distributions in the parameter space.  A Bayesian framework for VQS was introduced in \cite{jaakkola2000bayesian}, with the aim of enabling the estimation of the target parameter using variational techniques. Bayesian inference is also at the core of works such as \cite{quek2021adaptive}, which address the different problem of quantum tomography. 

Adaptive protocols play an important role in quantum computing and metrology. For example, in \cite{Demkowicz-Dobrzanski2017}, the authors present a foundational framework for estimating Hamiltonian parameters in the presence of general Markovian noise, establishing algebraic criteria that bound precision scaling under any adaptive strategy and linking these bounds to quantum error‑correction techniques.  Recently, the work \cite{nikoloska2025dynamic} has presented an adaptive learning strategy for VQS that adopts a frequentist design objective together with conformal inference. The method was shown to provide deterministic worst-case guarantees in terms of coverage for set prediction.

Bayesian inference is particularly useful for the design of adaptive protocols. In fact, uncertainty metrics can guide the exploration of the parameter space,  seeking to collect information that maximally reduces the epistemic uncertainty of the learner. Specifically, Bayesian experimental design is a statistical framework that uses current beliefs to plan experiments in a way that maximizes the expected information obtained from the results. This principle has been applied to quantum information tasks such as tomography \cite{von2014bayesian, kleinegesse2020bayesian, hickman2022bayesian}.  For instance, the authors in \cite{Valeri2023} demonstrate an on‑chip adaptive Bayesian protocol to estimate three optical phases simultaneously, showing the potential of adaptive techniques for resource-limited probes.




Based on this review, to the best of our knowledge, there is currently no adaptive Bayesian mechanism for VQS, which is the focus of this work.



\subsection{Main Contributions}
In this work, we investigate adaptive, online VQS protocols that adopt an information-seeking,  Bayesian approach. The main contributions are as follows.

\noindent \textbf{Adaptive Bayesian single-shot quantum sensing:} We introduce an adaptive Bayesian VQS framework in which the  quantum sensor acts  as an experimenter, using its beliefs about the parameter of interest to design a sensing policy with high epistemic value, effectively reducing the agent's uncertainty as shown in Fig. \ref{fig:sensing_scheme_indirect} (right). The agent's belief about the parameter of interest is based on a  world model,  learned online, for look-ahead sensing action planning \cite{hafner2025mastering, seo2023masked, van2024active}.  Optimal planning leverages the active information gain as the guiding principle \cite{van2024active, buckley2017free}.

\noindent \textbf{Extension to quantum sensor fusion:} We extend the proposed technique to multi-agent settings in which multiple sensors estimate the parameter of interest.  Each agent  maintains its own world model, pursuing an epistemic policy. By integrating complementary data via a principled Bayesian approach, the proposed sensor fusion strategy  reduces uncertainty and compensates for the limitations of individual sensors. 


\noindent \textbf{Experimental validation:} We demonstrate the effectiveness of the proposed dynamic estimation scheme on a quantum magnetometry case study. We consider both noise-free and noise-limited regimes, and we show that the proposed protocol can enable robust parameter estimation using a single probe.

The remainder of the paper is organized as follows. Section II formulates the problem of sequential quantum parameter estimation. Section III introduces the proposed adaptive Bayesian VQS approach, which is extended to multi-agent settings in Section IV. Section V presents numerical experiments for a magnetometry task, detailing the setup, benchmarks, and results. Section VI concludes the paper.

\section{Problem Formulation} \label{sec:qsens}
We consider the problem of sequentially estimating a target parameter of interest  using a single sensing probe. The real-valued target parameter $x_t$ is assumed to evolve over time steps indexed as $t=0, ..., T$  according to an unknown physical process. The samples $x_t$ are generally correlated across the time steps  $t=0, ..., T$. 
At time $t$, the target parameter $x_t$ determines the operation of a quantum channel $\mathcal{E} (x_t)$. 

As seen in Fig.~\ref{fig:sensing_scheme_indirect}, the quantum sensor generates a probe quantum state $\rho(\theta_t)$ to interact with the channel. Following the general VQS framework \cite{maclellan2024end}, the quantum state is prepared by a $n$-qubit quantum circuit  dependent on the variational parameters $\theta_t$.
Probing results in a perturbed quantum state $\rho(\theta_t, x_t)$, which is expressed as 
\begin{align}
    \rho (x_t, \theta_t)=\mathcal{E} (x_t) \rho(\theta_t),
\end{align} where the notation $\mathcal{E} (x_t) \rho(\theta_t)$ represents the post-channel state when the input state is $\rho(\theta_t)$ and the channel is  $\mathcal{E} (x_t)$.   The notation $ \rho (x_t, \theta_t)$ emphasizes the dependence of the perturbed probe on target parameter $x_t$ and variational parameters $\theta_t$.

The perturbed state $\rho (x_t, \theta_t)$ is then measured with a  positive operator-valued measurement (POVM) parameterized by a vector $\mu_t$. As in \cite{maclellan2024end}, we focus on local complete measurements in a rotated basis, which is realized by a layer of local rotation gates parameterized by $\mu_t$, followed by measurements in the local computational bases. The resulting measurement setting is denoted as  $\mathcal{M}_t = \{\Pi_m(\mu_t)\}_{m=1}^{2^n}$.  

The measurement $\mathcal{M}_t$ maps the perturbed state to a likelihood probability distribution $p(s_t \, | \, x_t, \theta_t, \mu_t)$ over measurement outcomes ${s}_t \in \{0,1\}^n$, which is given by\begin{align}\label{probs}
   s_t \sim p(s_t \, | \, x_t, \theta_t, \mu_t) = \text{Tr} \big[ \Pi_{{s}_t}(\mu_t) \rho(x_t, \theta_t) \big]. 
\end{align}
Using the observed data sample ${s}_t$, an estimate $\hat{x}_t$ of the target parameter $x_t$ is finally produced. The objective is to generate an estimate that closely approximates the true target $x_t$.

To guide the design of the sensing and measurement policy, we assume that, after producing the estimate $\hat{x}_t$, the sensor receives feedback about the true value $x_t$. In practice, this information can be obtained in controlled environments used for the calibration of the VQS system whereby training is carried out under known disturbances. Alternatively, the true value $x_t$ may be inferred from reward signals received upon acting on the estimate $x_t$. 




\section{Single-Shot Quantum Sensing}
In the proposed Bayesian VQS approach, the quantum sensor operates as an autonomous agent pursuing an information seeking policy. In particular, the agent uses a sensing action $a_t = \{\theta_t, \mu_t\}$, determining probe parameters $\theta_t$ and measurement parameters $\mu_t$ at time $t$, by maximizing the  \emph{active information gain} about the parameter of interest $x_t$ \cite{van2024active, haykin2012cognitive}. The executed action $a_t$ results in a new observation $s_t$ via \eqref{probs}, which then allows the agent to revisit its beliefs about future target parameters, guiding the selection of future actions.

As detailed next, the  update of the agent's belief and the evaluation of the active information gain rely on a {world model} -- in the sense of \cite{hafner2025mastering} -- capturing the agent's posterior belief on the parameters ${x}_t$, and on an {observation model} that allows the agent to generate hypothetical observations $\hat{s}_t$ for a given action $a_t$. 



\subsection{Probabilistic Modeling}

The agent maintains a joint distribution for the observations $s_{0:T}$ and for the parameters $x_{0:T}$, which is used to guide the selection future actions $a_t$. Denote as $\mathcal{D}_{0:t-1} = \{s_{0:t-1}, x_{0:t-1},a_{0:t-1}\}$ the dataset of previous observations $s_{0:t-1}$, parameters  ${x}_{0:t-1}$ and actions $a_{0:t-1}$ up to time $t-1$.  The joint distribution at hand is modeled via the factorization\begin{align}\label{mean_field}
     &q(s_{0:T}, x_{0:T} \, | \, \mathcal{D}_{0:T})  = \prod_{t=0}^T q(s_t, x_t \, | \, a_t, \mathcal{D}_{0:t-1}),
\end{align}
where the per-time step distribution $q(s_t, x_t \, | \, a_t, \mathcal{D}_{0:t-1})$ is given as
\begin{align}\label{gen}
     &q(s_t, x_t \, | \, a_t, \mathcal{D}_{0:t-1}) = q (s_t \, | \, x_t, a_t) q(x_t \, | \,  \mathcal{D}_{0:t-1}).
\end{align}In \eqref{gen},  the first term $q (s_t \, | \, x_t, a_t)$ represents an observation model, accounting for the likelihood of obtaining a particular observation $s_t$ given parameter $x_t$ and action $a_t$, whilst the second term, $q(x_t \, | \,\mathcal{D}_{0:t-1})$, models the agent's posterior belief about the  parameter $x_t$ given the available observations $\mathcal{D}_{0:t-1}$. These models are discussed in detail next.

\subsection{World Model}

The agent's posterior belief about the parameter $x_t$ is represented by the variational posterior distribution  $q(x_t \, | \,  \mathcal{D}_{0:t-1})$, which may be also referred to as world model \cite{hafner2025mastering}. Following standard modeling assumptions for low-dimensional data \cite{bishop2006pattern}, we adopt the parametric Gaussian model 
\begin{align}\label{world_sim}
q(x_t \, | \, \mathcal{D}_{0:t-1}) = \mathcal{N} (f_w(  \mathcal{D}_{0:t-1}), \sigma),
\end{align}
where the variational parameters $w$ determine the function $f_w(\mathcal{D}_{0:t-1})$ and $\sigma$ is the standard deviation. { The function $f_w( \mathcal{D}_{0:t-1})$ may be implemented using sequence models such as recurrent neural networks, state-space models, or transformers. } In more complex settings, with high-dimensional data, more powerful generative models, such as normalizing flows \cite{trippe2018conditional}, or diffusion models \cite{yang2023diffusion} can also be used. Alternatively, the posterior belief can also admit energy-based descriptions or joint-embedding architectures \cite{garrido2024learning}. 

\subsection{Observation Model}
To design an information seeking policy, the agent is assumed to have access to a 
simulator $\hat{\mathcal{E}} (x_{t})$ of the quantum channel $\mathcal{E} (x_{t})$.
This supports the generation of hypothetical measurements via the observation model
\begin{align}\label{probs_sim}
    q(s_{t} \, | \, x_{t}, a_{t}) = \text{Tr} \big[ \Pi_{s_t}(\mu_{t}) \phi(x_{t}, \theta_{t}) \big], 
\end{align}
where
\begin{align}
     \phi (x_{t}, \theta_{t})=\hat{\mathcal{E}}(x_{t}) \phi(\theta_{t}).
\end{align} 

In practice, the surrogate channel $\hat{\mathcal{E}} (x_{t})$ may be implemented on a NISQ device in order to circumvent the exponential complexity of classical simulations. In particular, using Stinespring dilation theorem \cite{busch2016dilation}, quantum channels acting on $n$ qubits can be simulated using at most $3n$ qubits on a NISQ device. This is done by including at most $2n$ ancilla qubits, which evolve through  a unitary transformation jointly with the $n$ qubits of interest.


\subsection{Active Information Gain}\label{sec:AIC}
The  model (\ref{mean_field}) allows the agent to pursue a sensing policy with high epistemic value. To this end, the agent uses actions $a_t$ that maximize the \textit{active information gain}. This is  defined as the mutual information $\mathrm{I}_q ({s}_{t}; {x}_{t} \, | \, {a}_{{t}}, \mathcal{D}_{0:t-1})$ between the parameter of interest $x_t$ and the observation $s_t$ conditioned on the available observations $\mathcal{D}_{0:t-1}$. The notation $\mathrm{I}_q(\cdot;\cdot)$ emphasizes that the mutual information is evaluated with respect to the variational joint distribution in (\ref{gen}).

The mutual information $\mathrm{I}_q ({s}_{t}; {x}_{t} \, | \, {a}_{{t}}, \mathcal{D}_{0:t-1})$ provides a measure of the reduction in uncertainty about the target parameter $x_t$ that  can be obtained by using action $a_{t}$ and collecting observation $s_t$ \cite{Simeone_2025}. In fact, it can be expressed as \begin{align}\label{info_gain}
    &\mathrm{I}_q ({s}_{t}; {x}_{t} \, | \, {a}_{{t}}, \mathcal{D}_{0:t-1}) \nonumber\\
    &= \mathrm{H}_q({x}_{t} \, | \, a_{t}, \mathcal{D}_{0:t-1}) - \mathrm{H}_q({x}_{t} \, | \, {s}_{t}, a_{t}, \mathcal{D}_{0:t-1})\nonumber\\
    &= \mathrm{H}_q({x}_{t} \, |  \mathcal{D}_{0:t-1}) - \mathrm{H}_q({x}_{t} \, | \, {s}_{t}, a_{t}, \mathcal{D}_{0:t-1}),
\end{align} where $\mathrm{H}_q(\cdot|\cdot)$ represents the conditional Shannon entropy and the second equality follows from the factorization in (\ref{gen}). 

Accordingly, the mutual information  $\mathrm{I}_q ({s}_{t}; {x}_{t} \, | \, {a}_{{t}}, \mathcal{D}_{0:t-1})$ equals the difference between the uncertainty  on the parameter $x_t$ given the available observations $\mathcal{D}_{0:t-1}$, accounted for by the entropy $\mathrm{H}_q({x}_{t} \, |  \mathcal{D}_{0:t-1})$, and the residual uncertainty on parameter $x_t$ after action $a_t$ is taken and an observation $s_t$ is collected, as quantified by the entropy $\mathrm{H}_q({x}_{t} \, | \, {s}_{t}, a_{t}, \mathcal{D}_{0:t-1})$. In words, it represents the reduction in uncertainty about parameter $x_t$ that is accrued thanks to action $a_t$. 

Overall, we propose to select the action that maximizes the active information gain, i.e.,  
\begin{align}\label{activeinfo}
    \underset{a_{t}}{\text{ max}} \,\,\,\,\,\, \mathrm{I}_q ({s}_{t}; {x}_{t} \, | \, {a}_{{t}}, \mathcal{D}_{0:t-1}).
\end{align}
This way, the agent selects actions that maximally reduce its own uncertainty about the parameter of interest $x_t$ according to its internal model (\ref{gen}). As explained above, the internal model (\ref{gen}) consists of a world model, $q(x_t \, | \,  \mathcal{D}_{0:t-1})$, as well as of an observation model, $q (s_t \, | \, x_t, a_t)$, with the latter enabling {active data generation} for the estimation of the information gain in the objective \eqref{activeinfo}.

As an implementation note, the optimization over the actions in (\ref{activeinfo}) is carried out using a single step of gradient descent, whereby the gradients are computed using parameter-shift rules \cite{schuld2018supervised,simeone2022introduction}. To this end, the mutual information in \eqref{info_gain}, and its gradient, are estimated using Monte Carlo sampling-based techniques  \cite{2020SciPy-NMeth} using the world and observation models in \eqref{world_sim} and \eqref{probs_sim}, respectively. 

\subsection{Estimator}
At each time step $t$, the estimate $\hat{x}_{t}$ is obtained by using the maximum a posteriori rule in Bayesian estimation \cite{kay1993fundamentals} under the model \eqref{gen}. Given the variational posterior \eqref{world_sim}, this   yields the posterior mean estimator $\hat{x}_{t}=f_w( \mathcal{D}_{0:t-1})$ which is the output of the world model. 


Having produced the estimate $\hat{x}_t$  and received feedback about the true parameter $x_t$, the agent revisits its beliefs about the parameter of interest, updates its world model in an online manner. This is done  using a single gradient descent step as\begin{align}\label{w_upd}
    w_{t+1} \leftarrow  w_{t} - \eta \nabla_w (x_t-f_w(\mathcal{D}_{t}))^2.
\end{align}

\section{Extension to Multiple Sensors}

In this section, we extend the approach presented in the previous section to multi-sensor systems.
In this setting, a collection of 
$K$ heterogeneous sensors collaboratively estimate a common target parameter $x_t$ via a fusion of the estimates obtained at each sensor. As detailed next, each sensor $k$ maintains a local variational model \eqref{gen}, takes actions $a^k_t=\{\theta^k_t,\mu^k_t\}$, and produces a local estimate  $\hat{x}^k_t$. The action consists of the local PQC parameters $\theta^k_t$ used by sensor $k$ and of the local POVM parameters $\mu^k_t$. The local estimates are then averaged to obtain a final global estimate $x_t$ at each time step $t$.

\subsection{Local World and Observation Models}
Each sensor collects its own datasets $\mathcal{D}^k_{0:t-1} = \{{s}^k_{0:t-1}, {x}_{0:t-1}, a^k_{0:t-1}\}$
of previous observations $s^k_{t-1}$, parameters $x_{0:t-1}$, and actions $a^k_{0:t-1}$ up to time $t-1$. Assuming the factorization in \eqref{mean_field}-\eqref{gen}, each sensor also independently maintains  per-time step variational distributions $q(s^k_t, x_t \, | \, a^k_t, \mathcal{D}^k_{0:t-1})=q (s^k_t \, | \, x_t, a^k_t) q(x_t \, | \,  \mathcal{D}^k_{0:t-1})$, encompassing the local world model
\begin{align}\label{probs_obs_1}
&q(x_t\, | \, \mathcal{D}^k_{0:t-1})= \mathcal{N} (f_{w^k}(\mathcal{D}^k_{0:t-1}),g_{w^k}(\mathcal{D}^k_{0:t-1}) )
\end{align}and  the observation model
\begin{align}\label{probs_sim1}
    q (s^k_t \, | \, x_t, a^k_t) = \text{Tr} \big[ \Pi_{s^k_t}(\mu^k_{t}) \phi(x^k_{t}, \theta^k_{t}) \big]. 
\end{align} Note that in (\ref{probs_obs_1}) we allow the standard deviation $\sigma^k_t=g_{w^k}(\mathcal{D}^k_{0:t-1})$ to vary across sensors $k$. This way, as discussed in the next subsection, the fusion of estimates obtained at different sensors can account for the relative estimated precisions of the respective estimates. 
Each agent uses the procedure in Section.~\ref{sec:AIC} and deploys the probe with parameters that maximize the information gain defined in \eqref{info_gain}. 

\subsection{Estimator via Sensor Fusion}

The estimates $\hat{x}^k_{t}=f_{w_t^k}( \mathcal{D}^k_{0:t-1})$ produced by all sensors $k=1,...,K$ are fused by weighted averaging, yielding the global estimates \cite{koliander2022fusion}
\begin{align}
    \hat{x}_{t}= \gamma \sum_{k=0}^{K} \frac{\hat{x}^k_{t}}{{\sigma}^2_k}
\end{align}
with $\gamma=( \sum_{k=1}^{K} 1/\sigma_k^2 )^{-1}. $  This corresponds to a minimum mean squared error estimate on the basis of independent observations with distributions given by \eqref{probs_obs_1}.

\section{Experiments}
In this section, we provide experimental results to validate the proposed quantum sensing scheme.

\subsection{Sensing Task}
\begin{figure}
    \centering
    \includegraphics[width = 0.5\textwidth]{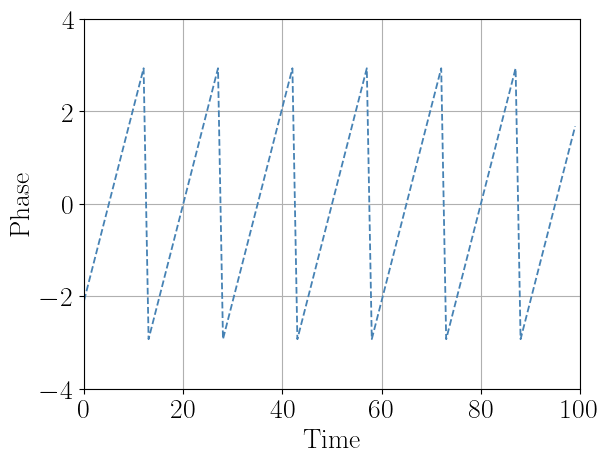}
    \caption{Target parameter $x_t$  as a function of time $t$. }
    \label{fig:phase}
\end{figure}

The parameter of interest $x_t$ is a phase evolving in a deterministic way according to the sawtooth process shown in Fig.~\ref{fig:phase}, i.e., 
\begin{align}
    x_t = 2 \pi \left( \left[\frac{t - t_0}{P}\right] - \left\lfloor{\frac{t - t_0}{P}}\right\rfloor \right), 
\end{align}
with $P = 15$s, over $100$ time steps. This setting may describe, e.g., a magnetometry setting in which the probe interacts with a magnetic field \cite{aiello2013composite, maayani2019distributed}.
The quantum channel is local and modeled as the tensor product of local channels applying separately to the $n$ qubits, i.e.,  $\mathcal{E} (x_t) =  R_z (x_t)^{\otimes n}$.

\begin{figure*}
    \centering
    \includegraphics[width = 0.9\textwidth]{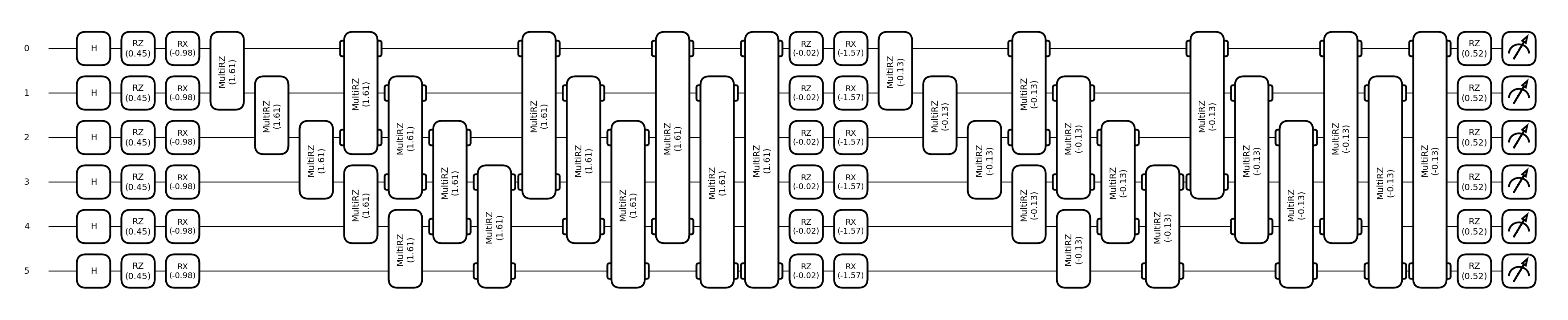}
    \caption{Probe-preparation ansatz used in the experiments. We consider an S(n)-equivariant ansatz consisting of single qubit gates and of parametric two-qubit gates applied in a cyclical manner across all pairs of successive qubits. We consider two layers and $n=6$ qubits.
    }
    \label{fig:ans_VQS}
\end{figure*}


\subsection{Probe-Preparation Ansatz and Hyperparameters}
We consider the S(n)-equivariant ansatz shown in Fig.~\ref{fig:ans_VQS}. This model  is equivalent to a fully connected quantum graph neural network \cite{mernyei2022equivariant}, which has been shown to be successful in learning useful meteorological probes \cite{verdon2019quantum}. The ansatz consists of general single qubit gates and of parametric two-qubit gates applied in a cyclical manner across all pairs of successive qubits. All single-qubit gates in the same layer share parameters, and so do all the two-qubit gates in the same layer.  We use two layers, and $n=6$ qubits. The vector $\mu$ parametrizes local Pauli Z rotation gates.

\subsection{World and Observation Models}
 The mean for the world model $f_w(\mathcal{D}_{0:t})$ is obtained as the output of a neural network parametrized by the parameters $w$. In a similar way, in the multi-agent setting, the mean $f_{w^k}(\mathcal{D}^k_{0:t})$ and the log-variance $g_{w^k}(\mathcal{D}^k_{0:t-1})$ are obtained using a neural network with parameters $w^k$ and two output neurons. The neural network is comprised of three hidden layers with $256$ hidden neurons each with ReLU activations. The input is given by the concatenated triplet $\{s_{t-1}, x_{t-1}, a_{t-1}\}$ from the previous step only. This reduces complexity, and it will be seen next to offer high performance even when compared to non-adaptive schemes based on more complex recurrent models.  The model is trained using Adam in an online manner, with a single gradient step at time $t$ and a learning rate of $0.001$. The model takes as input the past  The observation model simulates the channel using a unitary encoding, possibly followed by a parameter-independent quantum channel.

\subsection{Benchmarks}
We compare the proposed adaptive Bayesian technique against the following benchmarks:

\noindent \textbf{Static Bayesian scheme}: To investigate the effectiveness of the epistemic policy in the single-agent setting, as a benchmark, we consider a variational quantum sensing scheme that, whilst  learning a parametric estimator of the parameter of interest \cite{maclellan2024end, meyer2021variational, nikoloska2025dynamic}, uses randomly selected actions,  thereby  not pursuing an information-seeking policy. The estimator is an LSTM, an architecture specifically tailored for processing sequences, with two hidden layers with $256$ hidden neurons each and ReLu activations. We use L2 regularization and a multiplicative learning rate decay schedule, with decay of $0.1$ every $50$ time steps. 
Identically to the proposed protocol, the probe here is prepared by an S(n) equivariant ansatz and we measure local observables. 

\noindent \textbf{Adaptive Bayesian scheme with $K$ probes}: To investigate the benefits and limitations of sensor fusion as a benchmark, we consider the proposed adaptive Bayesian scheme using $K$ probes. Thereby, we compare a single agent using $K$ probes against $K$ agents using a single probe. Note that, the probe preparation overhead is identical for both schemes. 

\begin{figure}
    \centering
    \includegraphics[width = 0.5\textwidth]{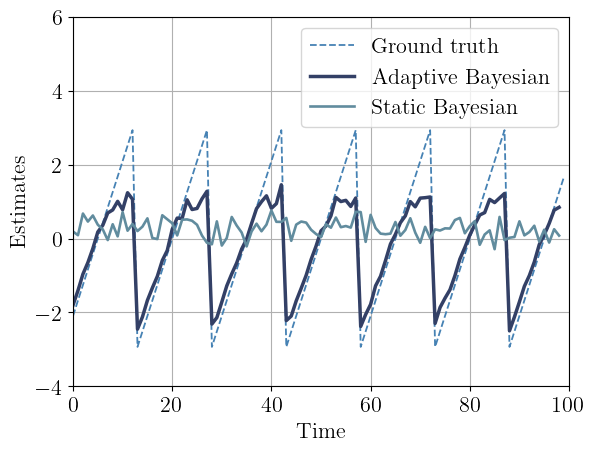}
    \caption{Phase estimates as a function of time. We consider a $n = 6$ qubit probe and an S(n) equivariant ansatz.}
    \label{fig:sensing_beliefs}
\end{figure}

\subsection{Results}

\noindent \textbf{Estimation capacity:} The estimated phase values as a function of time are shown in Fig.~\ref{fig:sensing_beliefs}. The proposed strategy outperforms the static Bayesian scheme, which produces estimates that significantly  differ from the target value. Conversely, the probe parameters that maximize information gain $\mathrm{I}_q (\hat{s}_t, \hat{x}_t \, | \, a_{t},  \mathcal{D}_{0:t-1})$, as selected by the proposed adaptive Bayesian scheme, can effectively enable accurate estimation of the target phase over time. This is a consequence of pursuing a sensing policy that helps the agent obtain its preferred data for learning a strong world model which, in turn, enables estimation.


\begin{figure}
    \centering
    \includegraphics[width = 0.5\textwidth]{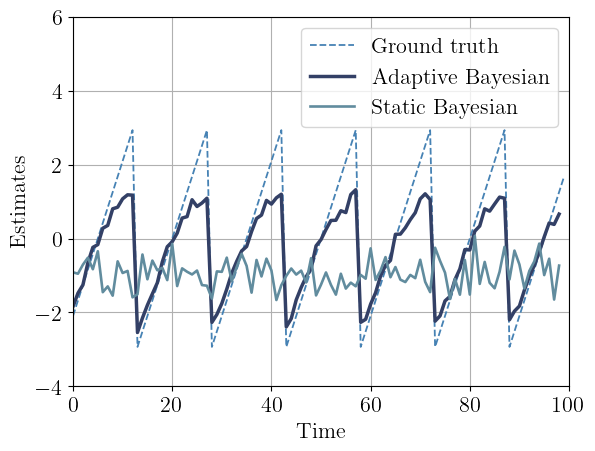}
    \caption{Phase estimates as a function of time in the presence of noise. The noise is sampled according to a Gaussian distribution $\mathcal{N}(0, 0.1)$. We consider a $n = 6$ qubit probe and an S(n) equivariant ansatz.}
    \label{fig:sensing_beliefs_noise}
\end{figure}
\noindent \textbf{Impact of noise:} Next, we investigate the impact of the noise present in near-term quantum (NISQ) devices. To this end, we add random noise, sampled according to a Gaussian distribution $\mathcal{N}(0, 0.1)$ to the parameters $\theta_t$ at each time step. This reflects, for instance, the fact that experimentalists can only manipulate the parameters with finite precision which is a realistic consideration in the NISQ era.
The estimated phase values as a function of time are shown in Fig.~\ref{fig:sensing_beliefs_noise}. The adaptive strategy performs well again, even in the presence of noise, and the probe parameters that maximize information gain can enable more accurate estimation of the  parameter of interest.

\begin{figure}
    \centering
    \includegraphics[width = 0.5\textwidth]{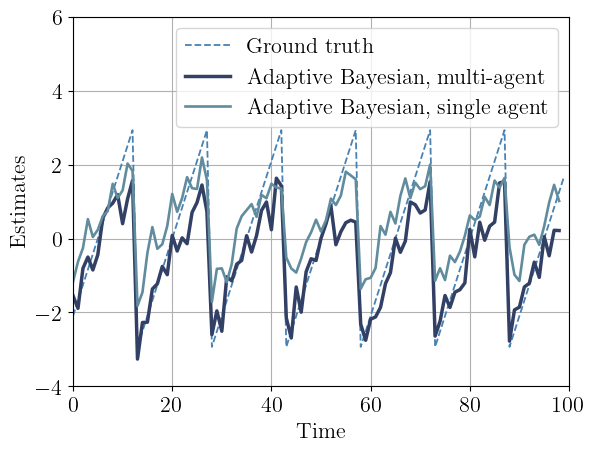}
    \caption{Phase estimates as a function of time in a multi-agent setting. We consider a system of $K=3$ agents, each equipped with a $n = 6$ qubit probe and an S(n) equivariant ansatz. As a benchmark, we use a single agent pursuing an information maximisation policy, using $3$ probes, and fusing the individual estimates.}
    \label{fig:sensing_beliefs_multi}
\end{figure}
\noindent \textbf{Benefits of quantum sensor fusion:} The results for the multi-sensor setting are shown in Fig.~\ref{fig:sensing_beliefs_multi}. We consider a system of $K=3$ agents. Each agent is equipped with a $n = 6$ qubit probe prepared using an S(n) equivariant ansatz. 
As it can be seen, when multiple agents are available, each attempting to maximize its information gain, one can obtain slightly more precise estimates compared to a single agent setup as a result of having diversity in the world models. However, this increases the probe preparation overhead by a factor of $K$.

\begin{figure}
    \centering
    \includegraphics[width = 0.5\textwidth]{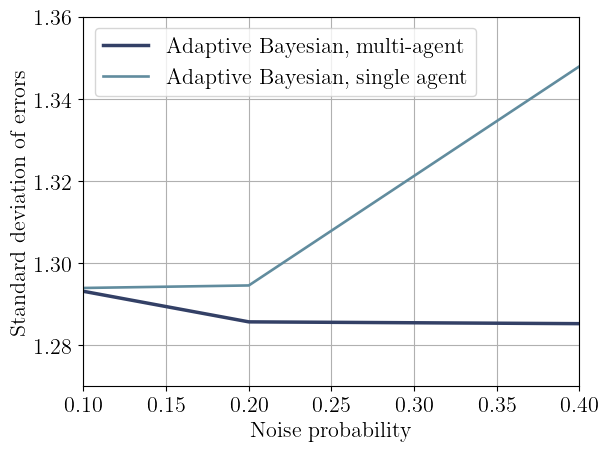}
    \caption{Standard deviation of the estimation errors $(x_t - \hat{x}_t)$ for different values of the noise probability. We consider a system of $K=3$ agents, each equipped with a $n = 6$ qubit probe and an S(n) equivariant ansatz. As a benchmark, we use a single agent pursuing an information maximisation policy, using $3$ probes, and fusing the individual estimates.}
    \label{fig:sensing_dev_multi}
\end{figure}
To investigate this further, in Fig.~\ref{fig:sensing_dev_multi} we show the standard deviation of the estimation errors $e_t = x_t - \hat{x}_t$, 
for different values of the noise probability. To this end, we add bit-flip errors (i.e., Pauli X gates) to each qubit in each layer of the circuit with probability $p \in \{0.1, 0.2, 0.4\}$. The multi-agent approach is seen to be more robust to increasing noise compared to the single-sensor system. This suggests that sensor fusion leads to more consistent and less variable estimates compared to the single-agent approach.

\section{Discussion and Conclusion}
In this work, we studied single-shot estimation of a target parameter using a quantum sensor. The quantum sensor, modeled as an agent learns a model of the world which captures the target evolution over time. We showed that using an information seeking policy with high epistemic value can enable learning precise world models. This, in turn, enables cheap, single-shot, target estimation. 

Various future research directions arise. 
The proposed protocol can admit a memory module, which may improve the sensing precession even further. In the multi-agent setting, the independent approach shown here can be replaced with more sophisticated policies where the agents collaboratively provide an estimate. In addition, the world model can admit more sophisticated architectures, as it is the case, for example, in the reinforcement learning literature. In this regard, it can also be leveraged for downstream tasks whereby the agent in addition to epistemic actions, can also take pragmatic actions to accomplish a higher-level goal. 

\section*{Acknowledgments}
The work of R. van Sloun was supported by the European Research Council (ERC) project US-ACT (101077368) 
The work of O. Simeone was supported by an Open Fellowship of the EPSRC (EP/W024101/1) and by the EPSRC project (EP/X011852/1).

\bibliographystyle{IEEEtran}
\bibliography{litdab.bib}

\end{document}